\documentclass[%
 twocolumn,
 amsmath,amssymb,
 aps,
 prd,
]{revtex4-2}

\usepackage{graphicx}% Include figure files
\usepackage{dcolumn}% Align table columns on decimal point
\usepackage{bm}% bold math
\usepackage[
colorlinks=true,
linkcolor=blue,
filecolor=blue,
citecolor=blue,
urlcolor=blue
]{hyperref}% add hypertext capabilities
\begin{document}

%\preprint{APS/123-QED}

%\title{Clock Noise Cancellation in the Fabry--Perot Type Space-Borne Gravitational-Wave Telescope}% Force line breaks with \\
\title{Clock Noise Cancellation in Heterodyne Links between Optical Cavities for Space-Borne Gravitational-Wave Telescopes}

\author{Yutaro Enomoto$^1$}
 \email{enomoto.yutaro@jaxa.jp}
\author{Subaru Shibai$^{1,2}$}
\author{Kiwamu Izumi$^1$}
\affiliation{$^1$Institute of Space and Astronautical Science, Japan Aerospace Exploration Agency,\\
3-1-1 Yoshinodai, Chuo-ku, Sagamihara City, Kanagawa 252-5210, Japan}
\affiliation{$^2$Department of Physics, The University of Tokyo,\\
7-3-1 Hongo, Bunkyo Ward, Tokyo 113-0033, Japan}%Lines break automatically or can be forced with \\
% \email{second.author@institution.edu}
%\affiliation{%

\date{\today}% It is always \today, today,

\begin{abstract}
Space-borne gravitational-wave telescopes are key to extend the observation band below $10\,\mathrm{Hz}$.
The use of inter-satellite optical cavities linked by heterodyne interferometry is a promising approach to reach the sensitivity level of $10^{-22}/\sqrt{\mathrm{Hz}}$ in the decihertz band.
While heterodyne interferometry is advantageous for relaxing arm-length control requirements, it introduces susceptibility to clock jitter, which can be a significant noise source.
In the back-linked Fabry--Perot (BLFP) interferometer aiming at the decihertz band, the required clock stability exceeds that of current space-qualified oscillators by more than an order of magnitude.
We propose a clock noise cancellation scheme that uses two heterodyne signals with positive and negative beat-note frequencies, naturally obtained using both incoming and outgoing laser beams of arm cavities without additional clock modulation schemes.
By forming a weighted combination of these signals with time-dependent coefficients, clock jitter contributions are eliminated while preserving gravitational-wave information.
We present the theoretical framework, analyze performance under realistic arm-length drifts, and validate the approach through time-domain simulations using parameters from the B-DECIGO concept.
Results show that the synthesized signal recovers the original sensitivity and even improves the signal-to-noise ratio by a factor of $\sqrt{2}$ for shot noise.

%RANDOM SENTENCES JUST TO FILL THE SPACE!
%For space-borne gravitational wave detectors, heterodyne interferometry is beneficial to avoid the need for mass actuation for precise length control of the interferometer arms.
%However, the use of heterodyne interferometry has some drawbacks compared to homodyne interferometry.
%One of the most significant issues is that phase measurements of the heterodyne beat-note signal are susceptible to clock jitter.
%In fact, in the case of a back-linked Fabry--Perot (BLFP) interferometer, achieving the sensitivity level of $\sim10^{-22}\,/\sqrt{\mathrm{Hz}}$ at $1\,\mathrm{Hz}$ requires clock stability of $\sim10^{-15}\,\mathrm{s}/\sqrt{\mathrm{Hz}}$ at $1\,\mathrm{Hz}$ assuming a heterodyne frequency of $\sim10\,\mathrm{MHz}$, which is more than one order of magnitude better than the stability of the best space-qualified ultra-stable oscillators (USOs).

%ANOTHER RANDOM SENTENCES JUST TO FILL THE SPACE!
\end{abstract}

\maketitle

%\tableofcontents

\section{\label{sec:intro}Introduction}
Observations of gravitational waves by terrestrial detectors have opened a new window to explore the Universe and astrophysics \cite{abbott2016b,abbott2017d}.
To extend observations to frequencies approximately below $10\,\mathrm{Hz}$, space-borne gravitational wave telescopes are being studied, including LISA, TianQin, Taiji, DECIGO, BBO, and TianGo \cite{amaro-seoane2017,luo2016,hu2017a,kawamura2006,harry2006,kuns2020}.
For space-borne gravitational wave telescopes, heterodyne interferometry is advantageous because it eliminates the need to precisely maintain the absolute arm lengths.
For this reason, most of the proposed missions employ heterodyne interferometry.
However, the use of heterodyne interferometry has some drawbacks compared to homodyne interferometry.
One of the most significant issues is that phase measurements of the heterodyne beat-note signal are susceptible to clock jitter.
In fact, in the case of a back-linked Fabry--Perot (BLFP) interferometer for decihertz band observation \cite{izumi2021a}, achieving the sensitivity at a level of $10^{-22}\,/\sqrt{\mathrm{Hz}}$ at $1\,\mathrm{Hz}$ requires clock jitter of approximately $10^{-15}\,\mathrm{s}/\sqrt{\mathrm{Hz}}$ or less at $1\,\mathrm{Hz}$ for a heterodyne frequency of $10\,\mathrm{MHz}$. This requirement calls for a clock that is better in the stability than the best space-qualified ultra-stable oscillators (USOs) \cite{yamamoto2023} by more than one order of magnitude.

In the context of LISA-like missions \cite{amaro-seoane2017,luo2016,hu2017a}, which employ optical transponders for inter-satellite laser links, constant efforts have been made to mitigate clock noise.
The pioneering work by Hellings \cite{hellings2001} pointed out using lasers with two frequencies is key to cancel clock noise.
The time-domain algorithms for clock noise cancellation in the time-delay interferometry were proposed theoretically \cite{tinto2018,hartwig2021} and studied experimentally \cite{yamamoto2022,xie2023,zeng2023}.
Similar approach has also been investigated for the arm locking scheme \cite{xu2024,xia2025}.
In these studies, local clocks on each spacecraft are up-converted to gigahertz range and transmitted to another spacecraft in the form of phase-modulation sidebands on the laser beams.
However, these methods cannot be directly applied to the BLFP interferometer since it employs Fabry--Perot cavities for inter-satellite laser links.

In this paper, we present a method to cancel clock noise in the BLFP interferometer by combining two heterodyne signals with positive and negative beat-note frequencies, which can be naturally obtained utilizing both of the incoming and outgoing laser beams in a single arm cavity without the need for phase modulation by up-converted clocks.
This paper describes the configuration and working principle of the clock noise cancellation scheme, along with the expected sensitivity of a BLFP interferometer with this method.
The numerical simulation results are also presented to validate the working principle even with slowly varying beat-note frequencies due to arm flexing in realistic heliocentric orbits.
The results show that the clock noise can be suppressed to a level below the sensitivity of the original proposal of the BLFP interferometer \cite{izumi2021a}.
This method resolves one of the fundamental challenges in realizing the strain sensitivity of $10^{-22}/\sqrt{\mathrm{Hz}}$ or better around $1\,\mathrm{Hz}$ with heterodyne links between inter-satellite optical cavities in the BLFP interferometer.

%a BLFP interferometer, and at the same time, provides a solution for clock noise in generic heterodyne interferometry with optical cavities.

\section{\label{sec:principle}Working principle}
In this section, we present the working principle of the clock noise cancellation method proposed for the BLFP interferometer.
After outlining the interferometer configuration, we derive the behavior of lasers locked to Fabry–Perot arm cavities with linear arm-length drifts.
We then formulate the heterodyne beat-note signals, discuss the impact of clock jitter, and show how a suitable linear combination of two beat-note signals achieves cancellation of clock noise while preserving the gravitational-wave signals.

\begin{figure}[ht]
  \centering
  \includegraphics[width=0.4\textwidth]{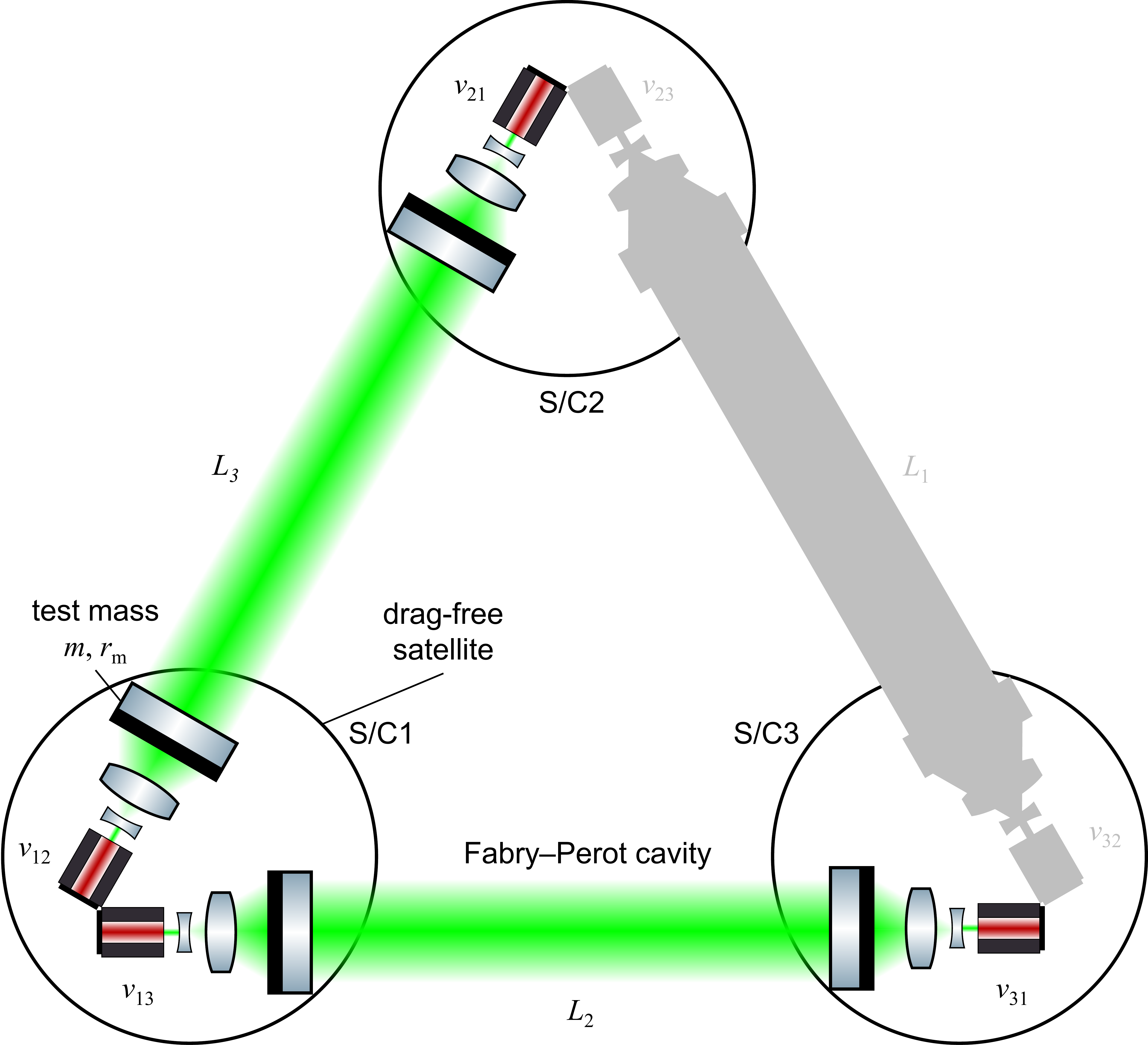}
  \caption{Conceptual configuration of a BLFP interferometer.
  The only components relevant to the signals obtained at S/C 1 are shown in color.}
  \label{fig:concept}
\end{figure}
\begin{figure}[ht]
  \centering
  \includegraphics[width=0.4\textwidth]{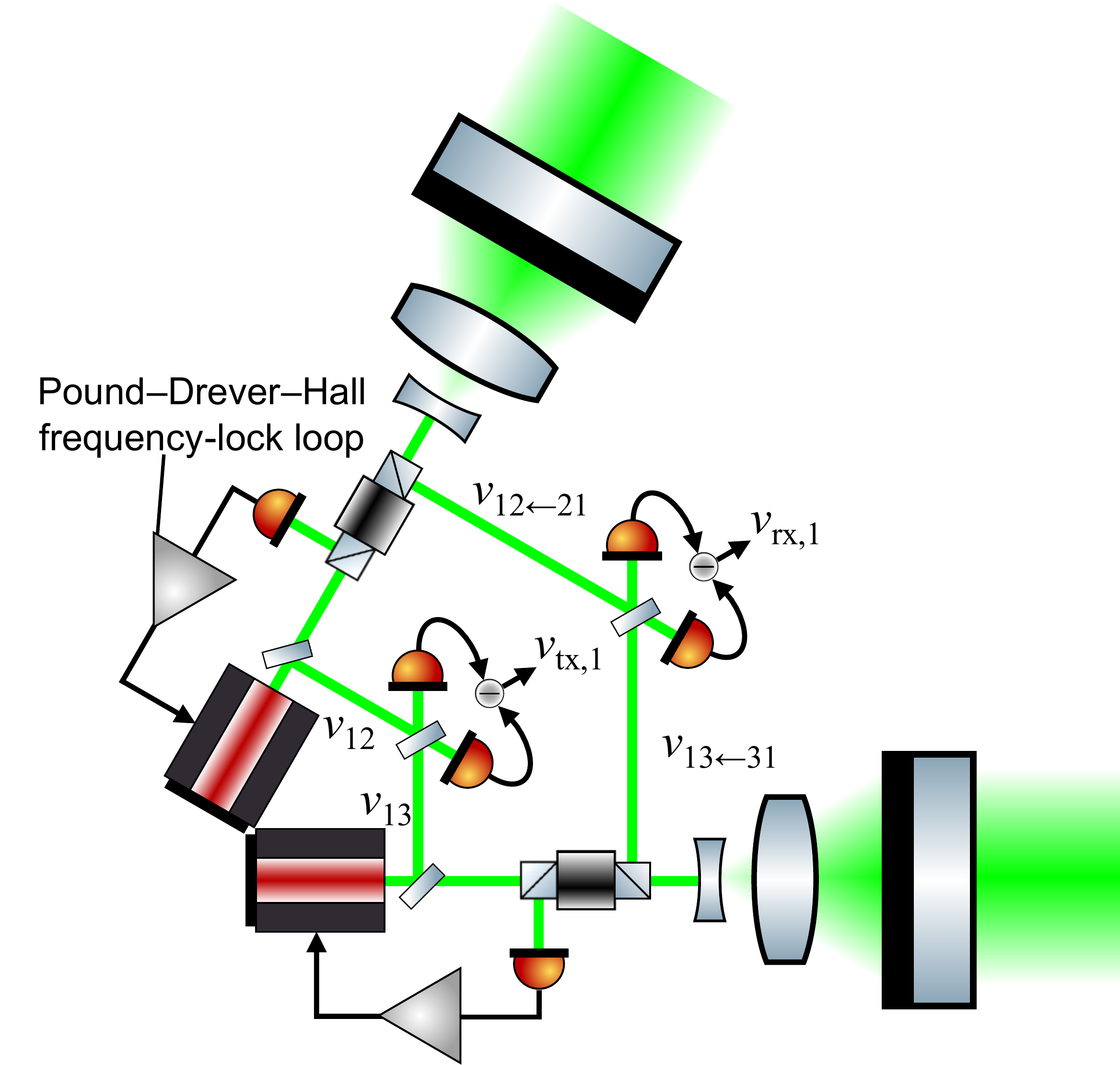}
  \caption{Closer look at the signals obtained at S/C 1.
  Two beat-note signals are obtained from the interference between incoming beams and outgoing beams.}
  \label{fig:two-beats}
\end{figure}

\subsection{\label{sec:config}Configuration}
Figure \ref{fig:concept} shows the conceptual configuration of a BLFP-type gravitational-wave detector.
In the triangular constellation, each spacecraft has two test masses and two lasers.
Test masses pointing to the other two spacecraft are used as the end mirrors of the arm cavities, and thus the inter-spacecraft cavities are formed along the sides of the triangle.
Two laser beams enter each arm cavity from opposite ends with orthogonal polarizations, ensuring that the two laser fields do not interfere optically.
At each spacecraft, the Michelson-like gravitational-wave signals are obtained from the interference between two lasers that are frequency-stabilized to the two arm cavities.
Following the naming convention in \cite{hartwig2021}, we denote the spacecraft as S/C 1, 2, and 3, and the frequency of the laser on S/C $i$ pointing to S/C $j$ as $\nu_{ij}$.
In addition, the frequency of the laser on S/C $i$ received at S/C $j$ is denoted as $\nu_{ij\leftarrow ji}$.
In a BLFP interferometer, the laser frequency is locked to the arm cavity resonance using the Pound--Drever--Hall technique so that the arm length variation is imprinted on the laser frequency change.
Therefore, the heterodyne beat-note signal between the two lasers following the two arms contains the information of the differential arm-length variation and thus the gravitational wave signal.

However, the heterodyne beat-note signal from two lasers with different frequencies is susceptible to clock jitter.
To mitigate the clock noise, we propose to utilize both of the incoming and outgoing laser beams in a single arm cavity to obtain two heterodyne beat-note signals.
Figure \ref{fig:two-beats} provides a more detailed view of how the beat-note signals are obtained at S/C 1.
The key difference from the original proposal is the acquisition of the second beat-note signal, which enables the cancellation of the clock noise.
The first beat-note signal $\nu_\mathrm{tx,1}$ is obtained from the interference between the outgoing beams from S/C 1, while the second beat-note signal $\nu_\mathrm{rx,1}$ is obtained from the interference between the incoming beams from S/C 2 and 3.
By appropriately setting the laser frequencies so that the beat-note frequencies of $\nu_\mathrm{tx,1}$ and $\nu_\mathrm{rx,1}$ have similar magnitudes but opposite signs, the clock jitter contributes differentially to them while gravitational waves contribute commonly.

\subsection{\label{sec:cavity}Fabry--Perot cavity with linear arm-length drift}
\begin{figure}
  \centering
  \includegraphics[width=0.4\textwidth]{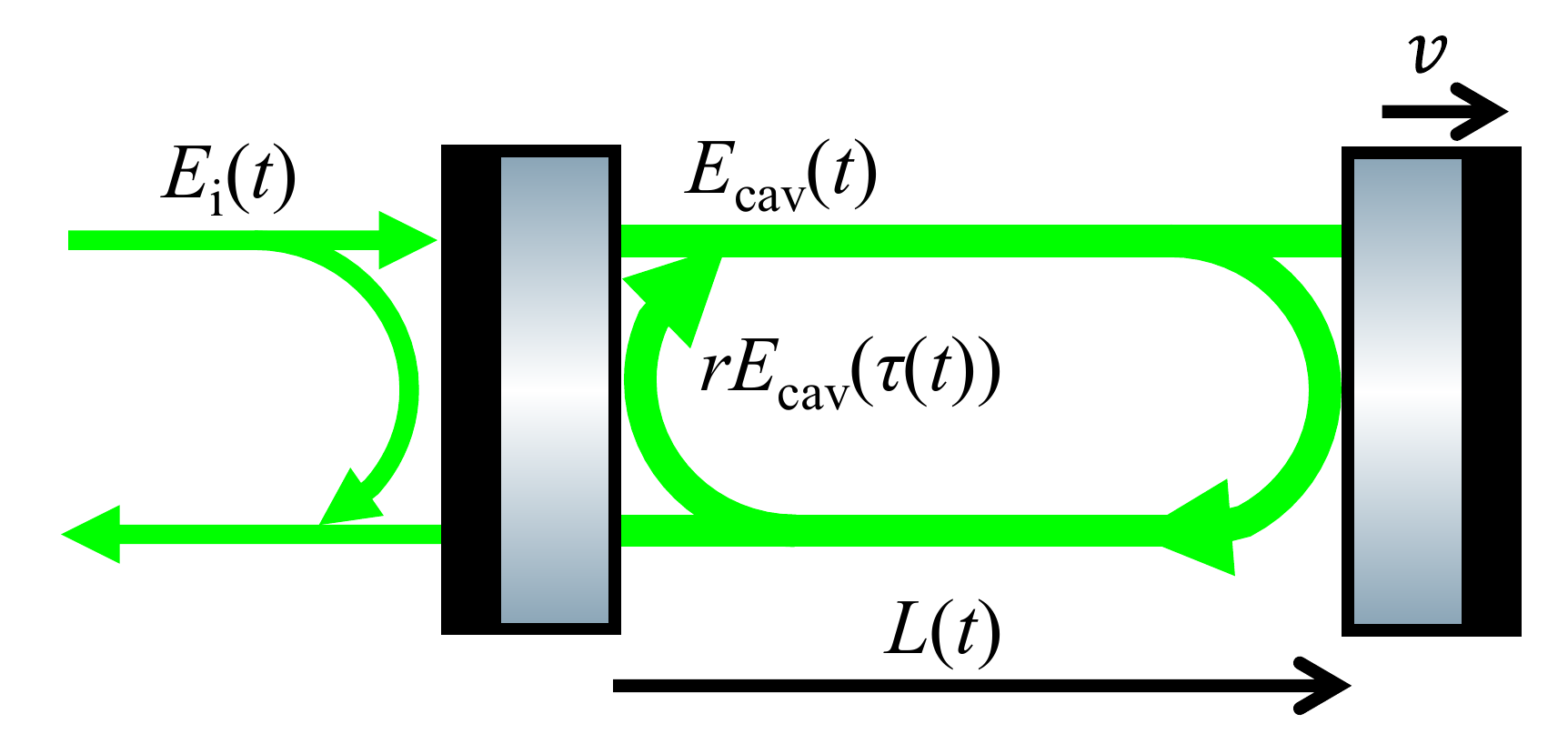}
  \caption{Fabry--Perot cavity with linear arm-length drift.
  The arm length is assumed to vary in time with a linear drift and small fluctuations.}
  \label{fig:FP}
\end{figure}
We first describe the frequency of the laser locked to a Fabry--Perot cavity.
Let us assume that the arm length change is the combination of a linear drift and small fluctuations so that the arm flexing is treated separately from the small fluctuations.
The arm length $L(t)$ is expressed as
\begin{align}
  L(t) &= L^o(t) + L^\epsilon(t),\\
  L^o(t) &= L_\mathrm{ini} + vt,
\end{align}
where $L_\mathrm{ini}$ is a constant representing the initial arm length, $v$ is also a constant for the drift velocity, and $L^\epsilon(t)$ is the small fluctuation.
Let $E_\mathrm{i}(t)$ and $E_\mathrm{cav}(t)$ be the incident electric fields at the input mirror of the cavity and the electric field inside the cavity, respectively (Fig. \ref{fig:FP}).
We rewrite the electric fields as
\begin{align}
  E_\mathrm{i}(t) &= A_\mathrm{i}e^{i\left[\phi^o(t) + \phi^\epsilon_\mathrm{i}(t)\right]},\\
  E_\mathrm{cav}(t) &= A_\mathrm{cav}e^{i\left[\phi^o(t) + \phi^\epsilon_\mathrm{cav}(t)\right]},\\
  \phi^o(t) &= \frac{\omega_\mathrm{ini}L_\mathrm{ini}}{v}\log\left(1 + \frac{vt}{L_\mathrm{ini}}\right).
\end{align}
Note that the instantaneous angular frequency $\omega^o(t):=d\phi^o(t)/dt$ is inversely proportional to the drifting arm length; $\omega(t) = L_\mathrm{ini}\omega_\mathrm{ini}/(L_\mathrm{ini} + vt)$.
For later convenience, we also define the instantaneous frequency $\nu(t) = \omega(t)/2\pi$ and its initial value $\nu_\mathrm{ini} \equiv \omega_\mathrm{ini}/2\pi$.

On one hand, from the superposition of the electric fields at the input mirror, the relation between the incident and intracavity fields is given by
\begin{align}
  E_\mathrm{cav}(t) &= \sqrt{1-r_\mathrm{m}^2}E_\mathrm{i}(t) + rE_\mathrm{cav}(\tau(t)),\label{eq:E_cav}
\end{align}
where $r_\mathrm{m}$ is the amplitude reflectivity of the input mirror, $r$ is the round-trip amplitude reflectivity of the cavity.
$\tau(t)$ is the retarded time defined as
\begin{align}
  \tau(t) &= \tau^o(t) + \tau^\epsilon(t),\\
  \tau^o(t) &= \frac{1-v/c}{1+v/c}t-\frac{2L_\mathrm{ini}/c}{1+v/c},
\end{align}
where $\tau^o(t)$ is the retarded time for the linear drift and $\tau^\epsilon(t)$ is the small fluctuation of the arm length including the effect of gravitational waves as well as force noise acting on the test masses.
In the low frequency regime where the typical frequency of the fluctuations is much lower than the free spectral range of the cavity $c/2L(t)$, $\tau^\epsilon(t)$ is expressed as $\tau^\epsilon(t) \simeq -2L^\epsilon(t)/c$.
Expanding Eq. (\ref{eq:E_cav}) and separating the zeroth- and first-order terms of $\phi^\epsilon_\mathrm{i}$, $\phi^\epsilon_\mathrm{cav}$, and $\tau^\epsilon$, we obtain
\begin{align}
  A_\mathrm{cav} &= \sqrt{1-r_\mathrm{m}^2}A_\mathrm{i} + rA_\mathrm{cav},\label{eq:E_cav_0}\\
  iA_\mathrm{cav}\phi^\epsilon_\mathrm{cav}(t) &= i \sqrt{1-r_\mathrm{m}^2}A_\mathrm{i}\phi^\epsilon_\mathrm{i}(t) + irA_\mathrm{cav}\phi^\epsilon_\mathrm{cav} \nonumber \\
  &\times\left[\phi^o(\tau(t))-\phi^o(\tau^o(t))+\phi^\epsilon_\mathrm{cav}(\tau^o(t))\right],\label{eq:E_cav_1}
\end{align}
assuming the resonance condition $\phi^o(t) - \phi^o(\tau^o(t)) = (\omega_\mathrm{ini}L_\mathrm{ini}/v)\log[(1+v/c)/(1-v/c)] = 2n\pi$ with $n$ an integer.
Substituting Eq. (\ref{eq:E_cav_0}) into Eq. (\ref{eq:E_cav_1}), we have
\begin{align}
  \phi^\epsilon_\mathrm{cav}(t) &= (1-r)\phi^\epsilon_\mathrm{i}(t) \nonumber \\
  &+ r\left[\phi^o(\tau(t))-\phi^o(\tau^o(t))+\phi^\epsilon_\mathrm{cav}(\tau^o(t))\right]. \label{eq:phi_cav}
\end{align}

On the other hand, from the Pound--Drever-Hall locking condition that the phases of the incident and leakage fields are the same, we have another relation between $\phi^\epsilon_\mathrm{i}$ and $\phi^\epsilon_\mathrm{cav}$ as
%\textcolor{red}{\bf [KI comment] what does 'aligned' mean?}
\begin{equation}
  \phi^\epsilon_\mathrm{i}(t) = \phi^o(\tau(t))-\phi^o(\tau^o(t))+\phi^\epsilon_\mathrm{cav}(\tau^o(t)) + N(t), \label{eq:PDH}
\end{equation}
with $N(t)$ being sensing noise in the Pound--Drever--Hall signal.
Combining Eqs. (\ref{eq:phi_cav}) and (\ref{eq:PDH}), we obtain
\begin{align}
  \phi^\epsilon_\mathrm{i}(t) - \phi^\epsilon_\mathrm{i}(\tau^o(t)) &= \phi^o(\tau(t))-\phi^o(\tau^o(t)) + \mathcal{N}(t)\\
  &= \omega(\tau^o(t))\tau^\epsilon(t) + \mathcal{N}(t), \label{eq:phi_i}
\end{align}
where $\mathcal{N}(t) = N(t) - rN(\tau^o(t))$.
This equation connects the phase fluctuation of the incident field $\phi^\epsilon_\mathrm{i}(t)$ to the arm length fluctuation $\tau^\epsilon(t)$, which is in turn read out by the heterodyne beat-note signal with the laser for the other arm.

To grasp the physical interpretation of Eq. (\ref{eq:phi_i}), we transform this under the assumption that $v=0$ and the typical frequency of the fluctuations is much lower than the free spectral range of the cavity $c/(2L_\mathrm{ini})$, obtaining
\begin{equation}
  \nu^\epsilon(t):=\frac{1}{2\pi}\frac{d\phi^\epsilon_\mathrm{i}(t)}{dt} \simeq -\nu_\mathrm{ini}\frac{L^\epsilon(t)}{L_\mathrm{ini}} + \frac{c\mathcal{N}(t)}{4\pi L_\mathrm{ini}}. \label{eq:phi_i_simple}
\end{equation}
This expression indicates that the frequency fluctuation of the laser locked to the cavity is proportional to the sum of the arm length fluctuation $\tau^\epsilon(t)$ and the sensing noise $\mathcal{N}(t)$.

In a similar way to Eqs. (\ref{eq:phi_cav}) and (\ref{eq:PDH}), we can also derive the expression for the phase fluctuation of the laser field transferred from the far end of the cavity.
When we express the transferred electric field at the input side as
\begin{equation}
  E_\mathrm{t}(t) = A_\mathrm{t}e^{i\left[\phi^o(t) + \phi^\epsilon_\mathrm{\leftarrow}(t)\right]},
\end{equation}
the phase fluctuation $\phi^\epsilon_\mathrm{\leftarrow}(t)$ is given by
\begin{align}
  \phi^\epsilon_\mathrm{\leftarrow}(t) - \phi^\epsilon_\mathrm{\leftarrow}(\tau^o(t)) &= \phi^o(\tau(t))-\phi^o(\tau^o(t)) + \mathcal{N}_\leftarrow(t)\\
  &= \omega(\tau^o(t))\tau^\epsilon(t) + \mathcal{N}_\leftarrow(t), \label{eq:phi_incoming}
\end{align}
where $\mathcal{N}_\leftarrow(t)$ is the sensing noise associated with the incoming beam having the same power spectral density as $\mathcal{N}(t)$.
Refer to Appendix \ref{sec:incoming} for the detailed derivation.

\subsection{\label{sec:heterodyne}Heterodyne beat-note signals and clock noise}
To apply the preceding discussions to the BLFP interferometer, we now need to define the variables by adding subscripts indicating the specific lasers or arms to all of those defined in the previous subsection.
For example, we define $\nu_{13}(t)$ to denote the frequency of the laser on S/C 1 pointing to S/C 3, and $\nu_{13\leftarrow31}(t)$ to denote the frequency of the laser on S/C 3 received at S/C 1.
Similarly, the arm length and its drift velocity between S/C 1 and 3 are denoted as $L_2(t)$ and $v_2$, respectively, while those between S/C 1 and 2 are denoted as $L_3(t)$ and $v_3$.

The heterodyne beat-note signals obtained at S/C 1 are formally given by
\begin{align}
  \nu_\mathrm{tx,1}(t) &= \nu_{13}(t) - \nu_{12}(t) + \frac{N_\mathrm{tx,1}(t)}{2\pi},\label{eq:beat1}\\
  \nu_\mathrm{rx,1}(t) &= \nu_{13\leftarrow31}(t) - \nu_{12\leftarrow21}(t) + \frac{N_\mathrm{rx,1}(t)}{2\pi},\label{eq:beat2}
\end{align}
where $N_\mathrm{tx,1}(t)$ and $N_\mathrm{rx,1}(t)$ are the sensing noise in the beat-note detections.
We decompose the beat-note frequencies into the offset and the fluctuation terms as follows:
\begin{align}
  \nu_\mathrm{tx,1}(t) &= \nu^o_\mathrm{tx,1}(t) + \nu^\epsilon_\mathrm{tx,1}(t)\\
  &= \left[ \nu^o_{13}(t) - \nu^o_{12}(t) \right]
  + \left[ \nu^\epsilon_{13}(t) - \nu^\epsilon_{12}(t) + \frac{N_\mathrm{tx,1}(t)}{2\pi} \right],\label{eq:nu_tx}\\
  \nu_\mathrm{rx,1}(t) &= \nu^o_\mathrm{rx,1}(t) + \nu^\epsilon_\mathrm{rx,1}(t)\\
  &= \left[ \nu^o_{13\leftarrow31}(t) - \nu^o_{12\leftarrow21}(t) \right] \nonumber \\
  &+ \left[ \nu^\epsilon_{13\leftarrow31}(t) - \nu^\epsilon_{12\leftarrow21}(t) + \frac{N_\mathrm{rx,1}(t)}{2\pi} \right].\label{eq:nu_rx}
\end{align}
We note that the beat-note frequencies $\nu_\mathrm{tx,1}(t)$ and $\nu_\mathrm{rx,1}(t)$ can be positive or negative in our convention.
Even though it is not possible to determine the sign of the frequencies solely with the instantaneous phasemeter outputs for the beat-note signals, the sign of the frequencies can be determined with the information of the response to $\nu_{13}(t)$ and $\nu_{13\leftarrow31}(t)$ combined, for example.
In the same approximation regime as Eq. (\ref{eq:phi_i_simple}), the fluctuation terms are expressed as
\begin{widetext}
\begin{align}
  \nu^\epsilon_\mathrm{tx,1}(t) &\simeq -\nu_{\mathrm{ini},13}\frac{L^\epsilon_2(t)}{L_\mathrm{ini,2}} + \nu_{\mathrm{ini},12}\frac{L^\epsilon_3(t)}{L_\mathrm{ini,3}} + \frac{c\mathcal{N}_{13}(t)}{4\pi L_\mathrm{ini,2}} - \frac{c\mathcal{N}_{12}(t)}{4\pi L_\mathrm{ini,3}} + \frac{N_\mathrm{tx,1}(t)}{2\pi},\\
  \nu^\epsilon_\mathrm{rx,1}(t) &\simeq -\nu_{\mathrm{ini},13\leftarrow31}\frac{L^\epsilon_2(t)}{L_\mathrm{ini,2}} + \nu_{\mathrm{ini},12\leftarrow21}\frac{L^\epsilon_3(t)}{L_\mathrm{ini,3}} + \frac{c\mathcal{N}_{13\leftarrow31}(t)}{4\pi L_\mathrm{ini,2}} - \frac{c\mathcal{N}_{12\leftarrow21}(t)}{4\pi L_\mathrm{ini,3}} + \frac{N_\mathrm{rx,1}(t)}{2\pi},
\end{align}
\end{widetext}
where $L_\mathrm{ini,2}$ and $L_\mathrm{ini,3}$ are the initial lengths of the arms between S/C 1 and 3 and between S/C 1 and 2, respectively.
We apply additional approximations that the difference in the arm lengths and laser frequencies are negligible ($L_\mathrm{2,ini} \simeq L_\mathrm{3,ini} \simeq L_\mathrm{ini}$ and $\nu_{\mathrm{ini},12} \simeq \nu_{\mathrm{ini},13} \simeq \nu_{\mathrm{ini},12\leftarrow21} \simeq \nu_{\mathrm{ini},13\leftarrow31} \simeq \nu_\mathrm{ini}$), and also explicitly separate the length fluctuations into terms of gravitational waves and force noise as 
\begin{equation}
  L^\epsilon_2(t)-L^\epsilon_3(t)=L_\mathrm{ini}h(t)+L^\mathrm{disp}_2(t)-L^\mathrm{disp}_3(t),
\end{equation}
where $h(t)$ is the gravitational wave strain and $L^\mathrm{disp}_2(t)$ and $L^\mathrm{disp}_3(t)$ are the force noise contributions to the corresponding arms expressed as displacement noise.
With these preparations, the fluctuation terms are simplified to
\begin{align}
  \nu^\epsilon_\mathrm{tx,1}(t) &\simeq \nu_\mathrm{ini}\left[h(t)+\frac{L^\mathrm{disp}_2(t)-L^\mathrm{disp}_3(t)}{L_\mathrm{ini}}\right] \nonumber \\
  &+\frac{c\mathcal{N}_{13}(t)}{4\pi L_\mathrm{ini}}-\frac{c\mathcal{N}_{12}(t)}{4\pi L_\mathrm{ini}}+\frac{N_\mathrm{tx,1}(t)}{2\pi},\label{eq:nu_tx_simple}\\
  \nu^\epsilon_\mathrm{rx,1}(t) &\simeq \nu_\mathrm{ini}\left[h(t)+\frac{L^\mathrm{disp}_2(t)-L^\mathrm{disp}_3(t)}{L_\mathrm{ini}}\right] \nonumber \\
  &+\frac{c\mathcal{N}_{13\leftarrow31}(t)}{4\pi L_\mathrm{ini}}-\frac{c\mathcal{N}_{12\leftarrow21}(t)}{4\pi L_\mathrm{ini}}+\frac{N_\mathrm{rx,1}(t)}{2\pi}.\label{eq:nu_rx_simple}
\end{align}
These equations tell us the way the gravitational wave signals are imprinted on the beat-note signals as well as the sensing and force noises.

Under the presence of clock jitter, the recorded beat-note frequencies $\tilde{\nu}_\mathrm{tx,1}(t)$ and $\tilde{\nu}_\mathrm{rx,1}(t)$ deviate from the true values $\nu_\mathrm{tx,1}(t)$ and $\nu_\mathrm{rx,1}(t)$, so that their fluctuation terms $\tilde{\nu}^\epsilon_\mathrm{tx,1}(t)$ and $\tilde{\nu}^\epsilon_\mathrm{rx,1}(t)$ are affected by the clock jitter.
By denoting the jitter of the clock on S/C 1 as $q_1(t)$, the recorded fluctuations of the beat-note signals are expressed as \cite{hartwig2021}
\begin{align}
  \tilde{\nu}^\epsilon_\mathrm{tx,1}(t) &= \nu^\epsilon_\mathrm{tx,1}(t) - \dot{q}_1(t)\nu^o_\mathrm{tx,1}(t), \label{eq:jitter_Tx}\\
  \tilde{\nu}^\epsilon_\mathrm{rx,1}(t) &= \nu^\epsilon_\mathrm{rx,1}(t) - \dot{q}_1(t)\nu^o_\mathrm{rx,1}(t), \label{eq:jitter_Rx}
\end{align}
where $\dot{q}_1(t)$ denotes the time derivative of $q_1(t)$.
These equations indicate that the clock jitter contributes to the two beat-note signals in proportion to their offset frequencies $\nu^o_\mathrm{tx,1}(t)$ and $\nu^o_\mathrm{rx,1}(t)$, which can be either positive or negative in our convention, depending on the choice of the laser frequencies.

\subsection{\label{sec:cancellation}Clock noise cancellation}
From the two recorded beat-note signals $\tilde{\nu}_\mathrm{tx,1}(t)$ and $\tilde{\nu}_\mathrm{rx,1}(t)$, we can synthesize a new signal $\tilde{\nu}_\mathrm{sum,1}(t)$ as
\begin{align}
  \tilde{\nu}_\mathrm{sum,1}(t) :&= C_\mathrm{tx,1}(t)\tilde{\nu}_\mathrm{tx,1}(t) + C_\mathrm{rx,1}(t)\tilde{\nu}_\mathrm{rx,1}(t)\\
  &= C_\mathrm{tx,1}(t)\nu^o_\mathrm{tx,1}(t) + C_\mathrm{rx,1}(t) \nu^o_\mathrm{rx,1}(t) \nonumber \\
  &+ C_\mathrm{tx,1}(t)\tilde{\nu}^\epsilon_\mathrm{tx,1}(t) + C_\mathrm{rx,1}(t)\tilde{\nu}^\epsilon_\mathrm{rx,1}(t).
\end{align}
By separating the offset and fluctuation terms as $\tilde{\nu}_\mathrm{sum,1}(t) = \nu^o_\mathrm{sum,1}(t) + \tilde{\nu}^\epsilon_\mathrm{sum,1}(t)$, we have
\begin{align}
  \nu^o_\mathrm{sum,1}(t) &= C_\mathrm{tx,1}(t)\nu^o_\mathrm{tx,1}(t) + C_\mathrm{rx,1}(t) \nu^o_\mathrm{rx,1}(t),\\
  \tilde{\nu}^\epsilon_\mathrm{sum,1}(t) &= C_\mathrm{tx,1}(t)\tilde{\nu}^\epsilon_\mathrm{tx,1}(t) + C_\mathrm{rx,1}(t)\tilde{\nu}^\epsilon_\mathrm{rx,1}(t).
\end{align}
Inserting Eqs. (\ref{eq:jitter_Tx}) and (\ref{eq:jitter_Rx}) into the above equation, we obtain
\begin{align}
  \tilde{\nu}^\epsilon_\mathrm{sum,1}(t) &= C_\mathrm{tx,1}(t)\nu^\epsilon_\mathrm{tx,1}(t) + C_\mathrm{rx,1}(t)\nu^\epsilon_\mathrm{rx,1}(t) \nonumber \\
  &- \dot{q}_1(t)\nu^o_\mathrm{sum,1}(t). \label{eq:synthesized}
\end{align}
From Eq. (\ref{eq:synthesized}), we see that the clock noise term can be eliminated by choosing the coefficients $C_\mathrm{tx,1}(t)$ and $C_\mathrm{rx,1}(t)$ such that $\nu^o_\mathrm{sum,1}(t) = 0$.
In fact, by setting
\begin{align}
  C_\mathrm{tx,1}(t) &= -\frac{\nu^o_\mathrm{rx,1}(t)}{\nu^o_\mathrm{tx,1}(t)-\nu^o_\mathrm{rx,1}(t)},\label{eq:C_tx}\\
  C_\mathrm{rx,1}(t) &= \frac{\nu^o_\mathrm{tx,1}(t)}{\nu^o_\mathrm{tx,1}(t)-\nu^o_\mathrm{rx,1}(t)},\label{eq:C_rx}
\end{align}
we have $\nu^o_\mathrm{sum,1}(t) = 0$ and $C_\mathrm{tx,1}(t) + C_\mathrm{rx,1}(t) = 1$.
Thus, the synthesized signal $\tilde{\nu}_\mathrm{sum,1}(t)$ is free from clock noise while retaining the gravitational wave signals similarly to the beat-note signal in the original BLFP interferometer.
%This means that the synthesized signal $\tilde{\nu}_\mathrm{sum,1}(t)$ is free from the clock jitter and contains only the gravitational wave signal almost in the same manner as the original BLFP interferometer beat-note signal of $\tilde{\nu}^\epsilon_\mathrm{tx,1}(t)$ since $C_\mathrm{tx,1}(t) + C_\mathrm{rx,1}(t) = 1$.

For the choice of laser frequencies, if we set the frequencies of the lasers pointing counter-clockwise around $\nu_\mathrm{standard}$ and those pointing clockwise around $\nu_\mathrm{standard} + \nu_\mathrm{offset}$, we have $\nu^o_\mathrm{tx,1}(t) \sim -\nu_\mathrm{offset}$ and $\nu^o_\mathrm{rx,1}(t) \sim +\nu_\mathrm{offset}$, so that $C_\mathrm{tx,1}(t) \sim C_\mathrm{rx,1}(t) \sim 0.5$.
This means that the gravitational wave signal is almost equally contained in both of the two beat-note signals, and thus the signal-to-noise ratio of the synthesized signal $\tilde{\nu}_\mathrm{sum,1}(t)$ with respect to the sensing noise in the Pound--Drever--Hall loops and the beat-note detections is almost as double as that of the original BLFP interferometer beat-note signal $\tilde{\nu}^\epsilon_\mathrm{tx,1}(t)$.

\section{\label{sec:sensitivity}Sensitivity}
\subsection{\label{sec:simulation}Simulation conditions}

\begin{table}
  \centering
  \caption{Simulation parameters. As for the noise terms, the amplitude spectral densities are shown. The sensing noise terms are assumed to be white, while the displacement noise and clock jitter are given as functions of frequency by Eqs. (\ref{eq:displacement_noise}) and (\ref{eq:clock_stability}), respectively. PDH stands for Pound--Drever--Hall.}
 \begin{ruledtabular}
  \begin{tabular}{l c r}
    {\bf Symbol} & {\bf Description} & {\bf Value}\\
    \hline
     $L_\mathrm{ini,2}$& Initial arm length & $100,000-100\,\mathrm{m}$\\
    $L_\mathrm{ini,3}$  & Initial arm length & $100,000+100\,\mathrm{m}$\\
    $v_2$& Arm drift velocity  & $-5\,\mathrm{nm/s}$\\
    $v_3$ &Arm drift velocity & $10\,\mathrm{nm/s}$\\
    $P_\mathrm{i}$& Laser power  & $500\,\mathrm{mW}$\\
    $\lambda$& Laser wavelength  & $515\,\mathrm{nm}$\\
    $\nu_\mathrm{ini,13}$ & Laser frequency  & $c/\lambda$\\
    $\nu_\mathrm{ini,12}$ & Laser frequency & $c/\lambda + 15.019\,969\,\mathrm{MHz}$\\
    $\nu_\mathrm{ini,13\leftarrow31}$&Laser frequency  & $c/\lambda + 14.974\,648\,\mathrm{MHz}$\\
   $\nu_\mathrm{ini,12\leftarrow21}$& Laser frequency  & $c/\lambda + 0.000\,337\,\mathrm{MHz}$\\
    $N_{ij}(t)$ \\
    $N_{ij\leftarrow ji}(t)$& PDH sensing noise 
     & $6.2\times10^{-10}\,\mathrm{rad/\sqrt{Hz}}$\\
   $N_\mathrm{tx,1}(t)$ \\$N_\mathrm{rx,1}(t)$& Beat-note sensing noise & $5.1\times10^{-9}\,\mathrm{rad/\sqrt{Hz}}$\\
    $L^\mathrm{disp}_2(t)$\\$ L^\mathrm{disp}_3(t)$& Displacement noise  & $S_\mathrm{disp}$ cf. Eq. (\ref{eq:displacement_noise})\\
    $m_{}$ & Mirror mass  & $30\,\mathrm{kg}$\\
    $r_\mathrm{m}$ &Mirror amplitude reflectivity  & $0.9615$\\
    $q_1(t)$& Clock jitter  & $S_q$ cf. Eq. (\ref{eq:clock_stability})\\
  \end{tabular}
  \end{ruledtabular}
    \label{tab:sim-conditions}
\end{table}

Based on the working principle described in the previous section, we performed numerical simulations to validate the clock noise cancellation scheme.
The simulation parameters are summarized in Table \ref{tab:sim-conditions}.
These parameters match those in the original BLFP interferometer proposal \cite{izumi2021a}.
In particular, the main parameters such as the arm length, laser power and wavelength, and mirror mass reflectivity are chosen to be the same as those in a proposed space-borne gravitational-wave detector, B-DECIGO \cite{nakamura2016}: $L_i\simeq 100\,\mathrm{km}$, $P_\mathrm{i}=500\,\mathrm{mW}$, $\lambda=515\,\mathrm{nm}$, $m=30\,\mathrm{kg}$, and $r_\mathrm{m}=0.9615$.

As for the noise terms, the sensing noise in the Pound--Drever--Hall signals $N_{ij}(t)$ and $N_{ij\leftarrow ji}(t)$ is assumed to be white with an amplitude spectral density of $6.2\times10^{-10}\,\mathrm{rad/\sqrt{Hz}}$, representing shot noise with the laser power $P_\mathrm{i}$ and wavelength $\lambda$.
The sensing noise in the beat-note detections $N_\mathrm{tx,1}(t)$ and $N_\mathrm{rx,1}(t)$ are also assumed to be white with an amplitude spectral density of $5.1\times10^{-9}\,\mathrm{rad/\sqrt{Hz}}$, corresponding to shot noise with the optical power of $60\,\mathrm{mW}$ on the photodetector.
The magnitude of displacement noise $L^\mathrm{disp}_2(t)$ and $L^\mathrm{disp}_3(t)$ is based on the assumption of white force noise acting on the test masses.
We assume two types of force noise: one is residual force noise that independently acts on each test mass with amplitude spectral density $S^{1/2}_\mathrm{res} = 1.0\times10^{-16}\,\mathrm{N/\sqrt{Hz}}$ for each test mass, and the other is radiation pressure noise that differentially acts on two end test masses with amplitude spectral density $S^{1/2}_\mathrm{rp} = 1.7\times10^{-16}\,\mathrm{N/\sqrt{Hz}}$ \cite{izumi2021a}.
In total, the magnitude of the displacement noise is given by
\begin{align}
  S_\mathrm{disp}^{1/2}(f)&= \frac{1}{m_{}(2\pi f)^2}\sqrt{2S_\mathrm{res} + 4S_\mathrm{rp}}\\
  &= 3.8\times10^{-19}\left(\frac{1\,\mathrm{Hz}}{f}\right)^2\,\mathrm{m/\sqrt{Hz}}. \label{eq:displacement_noise}
\end{align}
To model the stability of a space-qualified atomic USO, we set the clock jitter $q_1(t)$ to be
\begin{align}
  &S^{1/2}_q(f) \nonumber \\
  &= \sqrt{10^{-32}+10^{-28}\cdot\left[\left(\frac{1\,\mathrm{Hz}}{f}\right)+ \left(\frac{1\,\mathrm{Hz}}{f}\right)^3\right]} \,\mathrm{s/\sqrt{Hz}}, \label{eq:clock_stability}
\end{align}
following the references \cite{yamamoto2023,lilley2021}.

\begin{figure}
  \centering
  \includegraphics[width=0.5\textwidth]{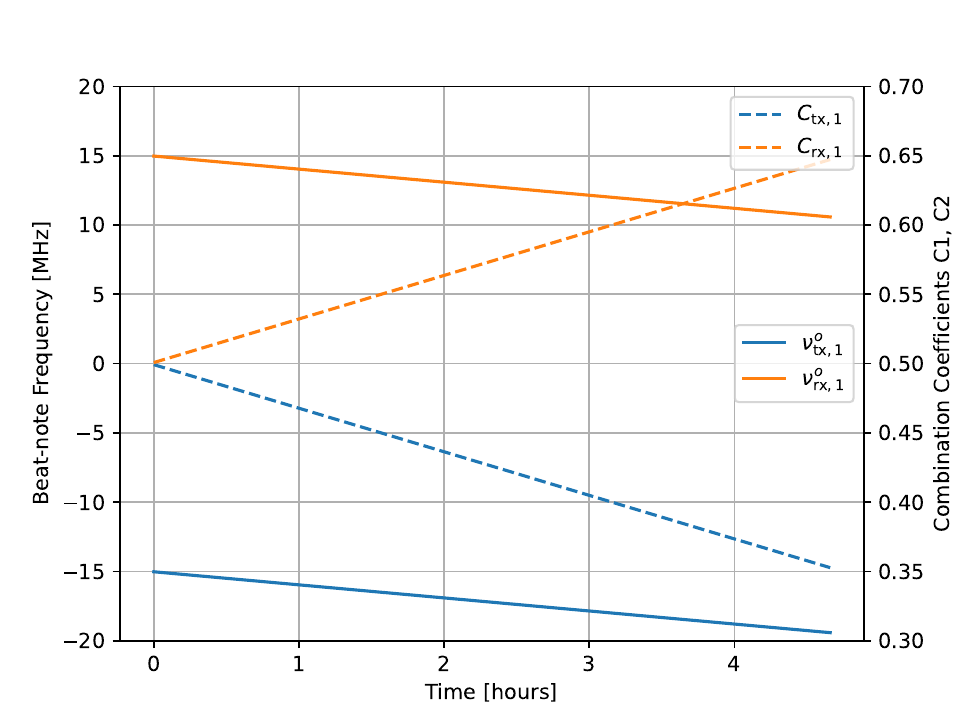}
  \caption{Simulated drift of the beat-note frequencies $\nu_\mathrm{tx,1}(t)$ and $\nu_\mathrm{rx,1}(t)$.
  The combination coefficients to cancel the clock noise $C_\mathrm{tx,1}(t)$ and $C_\mathrm{rx,1}(t)$ are also shown.}
  \label{fig:drift_combo}
\end{figure}

The magnitude of the arm length drift velocities is set to be an order of $10\,\mathrm{nm/s}$, to mimic the expected relative velocity of the spacecraft in a heliocentric orbit (Appendix \ref{sec:arm_flexing}).
With the chosen values of $v_2$ and $v_3$, the beat-note frequencies drift by approximately $1\,\mathrm{MHz}$ per hour (Fig. \ref{fig:drift_combo}).

The numerical simulation is performed in time domain with a sampling frequency of $1\,\mathrm{kHz}$ for a duration of approximately five hours.
The phase fluctuations of the incoming and outgoing laser beams on S/C 1 are calculated by solving Eqs. (\ref{eq:phi_i}) and (\ref{eq:phi_incoming}) with the Fourier transformation.
From the calculated phase fluctuations, $\nu^\epsilon_\mathrm{tx,1}(t)$ and $\nu^\epsilon_\mathrm{rx,1}(t)$ are calculated from the second terms of Eqs. (\ref{eq:nu_tx}) and (\ref{eq:nu_rx}).
The recorded phase fluctuations $\tilde{\nu}^\epsilon_\mathrm{tx,1}(t)$ and $\tilde{\nu}^\epsilon_\mathrm{rx,1}(t)$ under the presence of clock jitter $q_1(t)$ are calculated by Eqs. (\ref{eq:jitter_Tx}) and (\ref{eq:jitter_Rx}).
The jitter-free combination $\tilde{\nu}^\epsilon_\mathrm{sum,1}(t)$ is calculated by Eq. (\ref{eq:synthesized}) with the time varying coefficients $C_\mathrm{tx,1}(t)$ and $C_\mathrm{rx,1}(t)$ defined by Eqs. (\ref{eq:C_tx}) and (\ref{eq:C_rx}).

Figure \ref{fig:drift_combo} also shows the combination coefficients $C_\mathrm{tx,1}(t)$ and $C_\mathrm{rx,1}(t)$ that we use to cancel the clock noise.
The coefficients are determined by Eqs. (\ref{eq:C_tx}) and (\ref{eq:C_rx}) using the simulated beat-note frequencies $\nu^o_\mathrm{tx,1}(t)$ and $\nu^o_\mathrm{rx,1}(t)$.
We see that the coefficients vary slowly with time and deviate from $0.5$ as the beat-note frequencies drift.

\subsection{\label{sec:results}Results}
\begin{figure}[t]
  \centering
  \includegraphics[width=0.5\textwidth]{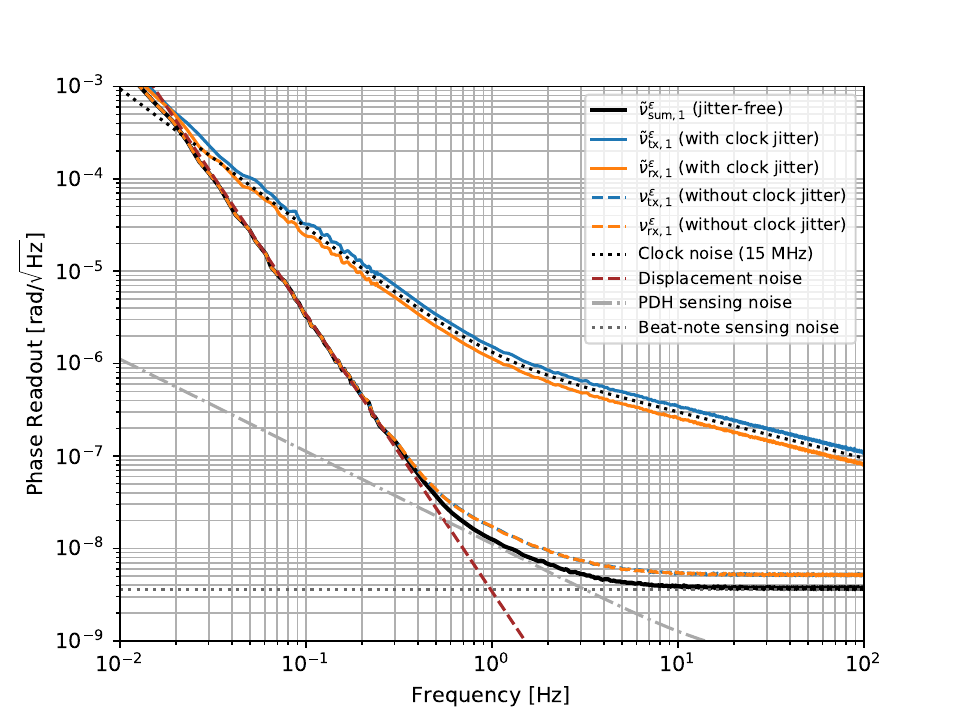}
  \caption{Simulated sensitivity curves of the beat-note signals $\tilde{\nu}_\mathrm{tx,1}(t)$ and $\tilde{\nu}_\mathrm{rx,1}(t)$ as well as the synthesized signal $\tilde{\nu}_\mathrm{sum,1}(t)$ in terms of phase noise.
  The estimated contributions from the sensing noise, displacement noise, and clock jitter are also shown.}
  \label{fig:phase_noise}
\end{figure}

\begin{figure}[ht]
  \centering
  \includegraphics[width=0.5\textwidth]{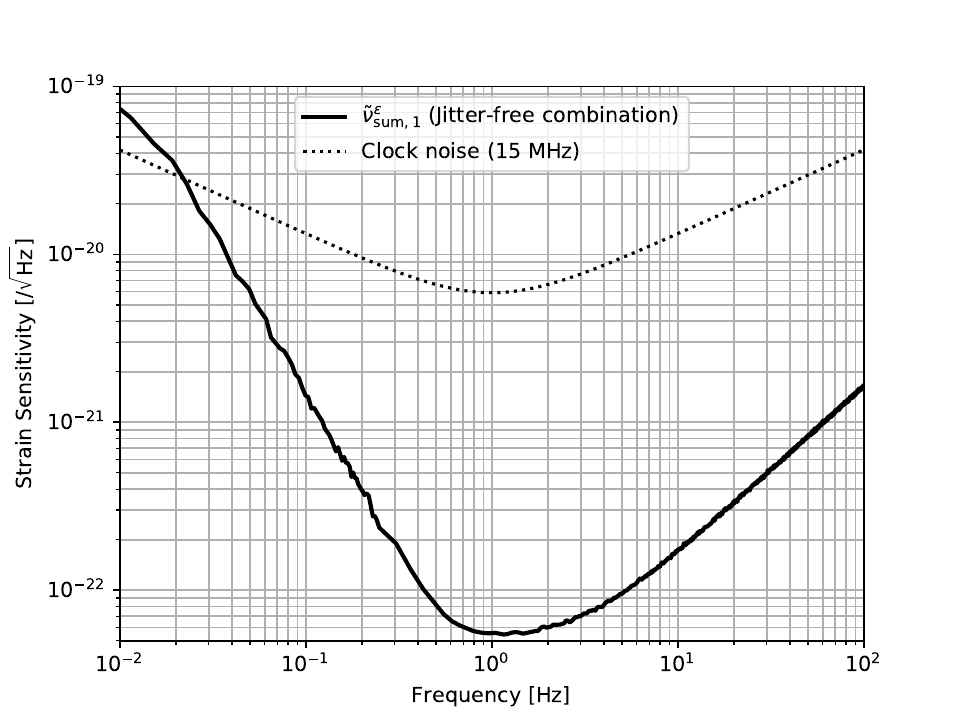}
  \caption{Simulated strain-equivalent sensitivity curve of the synthesized signal $\tilde{\nu}_\mathrm{sum,1}(t)$.
  The expected effect of the clock jitter if we do not apply the clock noise cancellation scheme is also shown.}
  \label{fig:strain_noise}
\end{figure}

Using the above parameters, we numerically simulated the time evolution of the beat-note signals $\tilde{\nu}_\mathrm{tx,1}(t)$ and $\tilde{\nu}_\mathrm{rx,1}(t)$ in time domain.
We also synthesized the clock-noise-free signal $\tilde{\nu}_\mathrm{sum,1}(t)$ from them.
Figure \ref{fig:phase_noise} shows the sensitivity curves of the three beat-note signals in terms of phase noise.
Without the clock jitter, the sensitivities of $\nu_\mathrm{tx,1}(t)$ and $\nu_\mathrm{rx,1}(t)$ are the same level as the original BLFP proposal.
However, with the clock jitter, the sensitivities of them are significantly degraded by more than one order of magnitude ($\tilde{\nu}_\mathrm{tx,1}(t)$ and $\tilde{\nu}_\mathrm{rx,1}(t)$).
The amount of degradation is consistent with the estimation of the clock jitter contribution with the beat-note frequencies assumed to be $15\,\mathrm{MHz}$ (black dotted line).

The black solid line in Fig. \ref{fig:phase_noise} shows the sensitivity of the synthesized signal $\tilde{\nu}_\mathrm{sum,1}(t)$.
The sensitivity is nearly restored to the jitter-free level and is slightly better above approximately $0.5\,\mathrm{Hz}$.
This reflects the fact that the gravitational wave signal is almost equally contained in both of the two beat-note signals while the PDH and beat-note sensing noises contained in them are statistically independent, so that the signal-to-noise ratio of the synthesized signal $\tilde{\nu}_\mathrm{sum,1}(t)$ with respect to the sensing noise is improved by a factor of $\sqrt{2}$.
This is consistent with the estimation of the sensing noise contributions (grey dashed and dotted lines) assuming that the sensing noise contributions in $\tilde{\nu}_\mathrm{tx,1}(t)$ and $\tilde{\nu}_\mathrm{rx,1}(t)$ are statistically independent.
The estimations of the noise elements including the displacement noise (brown dashed line) are derived from the approximated formula of Eqs. (\ref{eq:nu_tx_simple}) and (\ref{eq:nu_rx_simple}), which explains the simulation results well.
Figure \ref{fig:strain_noise} shows the strain-equivalent sensitivity curve of the synthesized signal $\tilde{\nu}_\mathrm{sum,1}(t)$ as well as the expected effect of the clock jitter if we do not apply the clock noise cancellation scheme.
The sky- and polarization averaging factor of $\sqrt{1/5}$ and the factor $\sqrt{3/4}$ coming from the fact that the two arms are not orthogonal but have an angle of $60^\circ$ are taken into account in the conversion from the phase readout to the strain.

\section{\label{sec:discussion}Discussion}
Even though the contribution of clock noise is eliminated by the proposed cancellation scheme, phase measurement in the sensitivity level of $10^{-8} \,\mathrm{rad/\sqrt{Hz}}$ or less is still demanding.
In fact, the noise level of the current state-of-the-art phase meters is from $\sim10^{-6}\,\mathrm{rad/\sqrt{Hz}}$ for the phasemeter based on all-digital phase-locked loop (ADPLL) \cite{bode2024} to $\sim10^{-7}\,\mathrm{rad/\sqrt{Hz}}$ for the phasemeter based on zero-crossing detection \cite{kokuyama2016} at $0.1\,\mathrm{Hz}$.
While there are technical noise sources in the digital domain that limit the current performance of the phasemeters as discussed in \cite{bode2024}, the fundamental noise sources in the ADPLL-based phasemeter are originated from the analog domain, which are quantization noise in the analog-to-digital converter and timing jitter of the clock.
As for the quantization noise, effective number of bits of $>16$ with sampling frequency of $100\,\mathrm{MHz}$ corresponds to the noise level of $<1.3\times10^{-9}\,\mathrm{rad/\sqrt{Hz}}$.
Therefore, it is important to eliminate the effect of timing jitter of the clock in the phasemeter.

In the simulation of this paper, we have assumed that the combination coefficients $C_\mathrm{tx,1}(t)$ and $C_\mathrm{rx,1}(t)$ are exactly known.
In practice, however, there will be some uncertainties in determining them due to the errors in measuring the beat-note frequencies.
Let us estimate the order of the necessary accuracy.
Since Fig. \ref{fig:phase_noise} shows that the contribution of the clock jitter is approximately two orders of magnitude larger than the target sensitivity level, the necessary accuracy of measuring the beat-note frequencies is estimated to be the order of $1\%$ or better, which is not challenging.
We have also assumed that the suppression of frequency noise by the PDH locking is sufficient.
If the suppression is insufficient, as is expected to be the case in reality, the residual frequency noise in the beat-note signals has to be removed in post-processing using the error signals of the PDH loops.

We note that shot noise in the heterodyne beat-note detection limits the sensitivity above approximately $3\,\mathrm{Hz}$ (Fig. \ref{fig:phase_noise}), and can be a bottleneck to further improve the sensitivity.
The shot noise level of the beat-note signals can be reduced by applying the squeezing technique of phase-insensitive heterodyne detection recently proposed and demonstrated in \cite{anai2024}.

This method of clock noise cancellation is potentially applicable to other precision measurements.
For example, the difference of the resonant frequencies of two optical cavities is measured by heterodyne interferometry in the experiments  measuring the thermal noise of cavity mirrors \cite{chalermsongsak2015,gras2018} and testing the Lorentz invariance of the speed of light \cite{eisele2009}.
%at the same time, provides a solution for clock noise in generic heterodyne interferometry with optical cavities.

%\begin{itemize}
%  \item Other noise sources in the phase meter (referring the context of modulation noise in LISA)
%  \item Necessary accuracy of $C_\mathrm{tx,1}$ and $C_\mathrm{rx,1}$
%  \item Insufficient frequency suppression by the PDH loop (which can hopefully be addressed in post-processing using the error signals of the PDH loops)
%  \item Possible reduction of the shot noise level in the beat-note signals applying the beat-note squeezing technique \cite{anai2024}
%\end{itemize}

\appendix

\section{\label{sec:incoming}Phase of the incoming beam}
\begin{figure}[ht]
  \centering
  \includegraphics[width=0.4\textwidth]{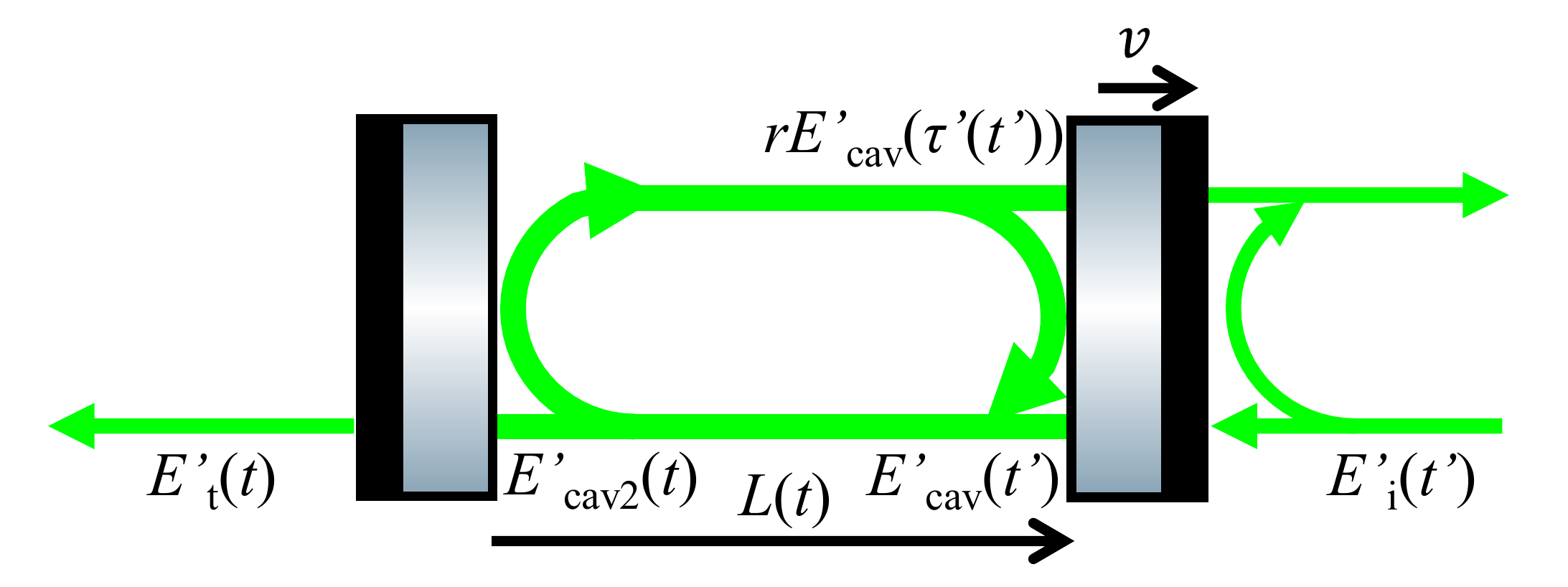}
  \caption{Schematic of the Fabry--Perot cavity and the tranferred beam from the far spacecraft.}
  \label{fig:incoming}
\end{figure}
In section \ref{sec:cavity}, we derive the expression of the phase fluctuation of the laser field locked to the cavity seen at the input side, which we call the outgoing beam.
Here, we derive the expression of the phase fluctuation of the laser field transferred from the far end of the cavity, which we call the incoming beam (Fig. \ref{fig:incoming}).

In the similar way to the derivation of Eq. (\ref{eq:phi_i}), the phase fluctuation of the laser field at the far end of the cavity can be derived.
By denoting the incident electric field at the far end and the electric field inside the cavity as
\begin{align}
  E'_\mathrm{i}(t') &= A'_\mathrm{i}e^{i\left[\phi^o(t') + \phi'^\epsilon_\mathrm{i}(t')\right]},\\
  E'_\mathrm{cav}(t') &= A'_\mathrm{cav}e^{i\left[\phi^o(t') + \phi'^\epsilon_\mathrm{cav}(t')\right]},
\end{align}
we express the phase fluctuations outside and inside the cavity as
\begin{align}
  \phi'^\epsilon_\mathrm{i}(t')-\phi'^\epsilon_\mathrm{i}(\tau'^o(t')) &= \phi^o(\tau'(t')) - \phi^o(\tau'^o(t')) + \mathcal{N}'(t'),\\
  \phi'^\epsilon_\mathrm{cav}(t')-\phi'^\epsilon_\mathrm{cav}(\tau'^o(t')) &= \phi^o(\tau'(t')) - \phi^o(\tau'^o(t')) \nonumber\\
  &+ (1-r)N'(t')
\end{align}
where the primed variables correspond to those defined for the incoming beam.
In particular, $t'$ is defined such that the incoming beam that departs at time $t'$ from the far end arrives at the near end at time $t$.
Thus $t'$ can be seen as a function of $t$ as
\begin{equation}
  t'(t) = t'^o(t) + t'^\epsilon(t),
\end{equation} 
where $t'^o(t) = (t+\tau^o(t))/2$.
Let us also define the electric field of the incoming beam inside the cavity at the near end by $E'_\mathrm{cav2}(t) = A'_\mathrm{cav}e^{i\left[\phi^o(t) + \phi'^\epsilon_\mathrm{cav2}(t)\right]}$.
Since $\phi'^\epsilon_\mathrm{cav2}(t) = \phi'^\epsilon_\mathrm{cav}(t')+\phi^o(t'(t))-\phi^o(t'^o(t))$ due to the retardation of the single trip, we have
\begin{align}
  \phi'^\epsilon_\mathrm{cav2}(t) - \phi'^\epsilon_\mathrm{cav2}(\tau(t)) &= \phi^o(\tau(t)) - \phi^o(\tau^o(t)) \nonumber \\
  &+ (1-r)N'(t').
\end{align}
Let us derive the phase fluctuation of the incoming beam seen at the near end, $E'_\mathrm{t}(t)=A'_\mathrm{t}e^{i\left[\phi^o(t) + \phi^\epsilon_\mathrm{\leftarrow}(t)\right]}$.
Considering that the vacuum field is also incident from the near end, we finally obtain the phase fluctuation of the incoming beam as
\begin{align}
  \phi^\epsilon_\mathrm{\leftarrow}(t) - \phi^\epsilon_\mathrm{\leftarrow}(\tau(t)) = \phi^o(\tau(t)) - \phi^o(\tau^o(t)) +\mathcal{N}_\leftarrow(t),\\
  \mathcal{N}_\leftarrow(t) = (1-r)N'(t') + N_\rightarrow(t)-N_\rightarrow(\tau(t)),
\end{align}
where $\phi^\epsilon_\mathrm{\leftarrow}(t)$ is the phase fluctuation of the incoming beam seen at the near end, and $N_\rightarrow(t)$ is the effect of the vacuum noise incident from the near end.
If we assume that the cavity is critically coupled, the magnitudes of $N'(t')$ and $N_\rightarrow(t)$ are the same.

Let us consider the power spectral densities of $\mathcal{N}(t)$ and $\mathcal{N}_\leftarrow(t)$, which we call $S_\mathcal{N}(f)$ and $S_{\mathcal{N}_\leftarrow}(f)$, respectively, under the assumption that $v=0$ and the typical frequency of the fluctuations is much lower than the free spectral range of the cavity $c/(2L_\mathrm{ini})$.
By denoting the power spectral densities of $N(t)$, $N'(t)$ and $N_\rightarrow(t')$ as $S_N$, we have
\begin{align}
  S_\mathcal{N}(f) &= (1-r)^2\left[1+(f/f_c)^2\right]S_N,\\
  S_{\mathcal{N}_\leftarrow}(f) &= (1-r)^2\left[1+(f/f_{c_\leftarrow})^2\right]S_N,
\end{align}
where $f_c = c(1-r)/(4\pi \sqrt{r} L_\mathrm{ini})$ and $f_{c_\leftarrow} = c(1-r)/(4\pi L_\mathrm{ini})$.
These corner frequencies are almost the same for $r\sim 1$, and thus the power spectral densities of the two noise terms are approximately equal as $S_{\mathcal{N}_\leftarrow}(f) \simeq S_\mathcal{N}(f)$.

\section{\label{sec:arm_flexing}Arm length drift}
When the three spacecraft are deployed into the cartwheel configuration in the heliocentric orbit, the inter-satellite distances are not constant but vary as a function of time due to the gravity from the celestial bodies whose primary contribution is that from the Sun.
Here, we estimate the typical drift velocity of the arm lengths or equivalently the inter-satellite distances following the reference \cite{nayak2006}. The tilt of the triangular formation plane with respect to the ecliptic plane is assumed to be a nominal value of $\pi/3$. Even though this setting for tilt is suboptimal in that a small tuning of the tile angle would allow for smaller variations, we choose this assumption for deriving the typical drift velocity. The numerical simulation does not include the gravitational influences from the Earth or other planets. It only incorporates the Sun only, which is sufficient for estimating the size of relative velocities.

\begin{figure}[ht]
    \centering
    \includegraphics[width=\columnwidth]{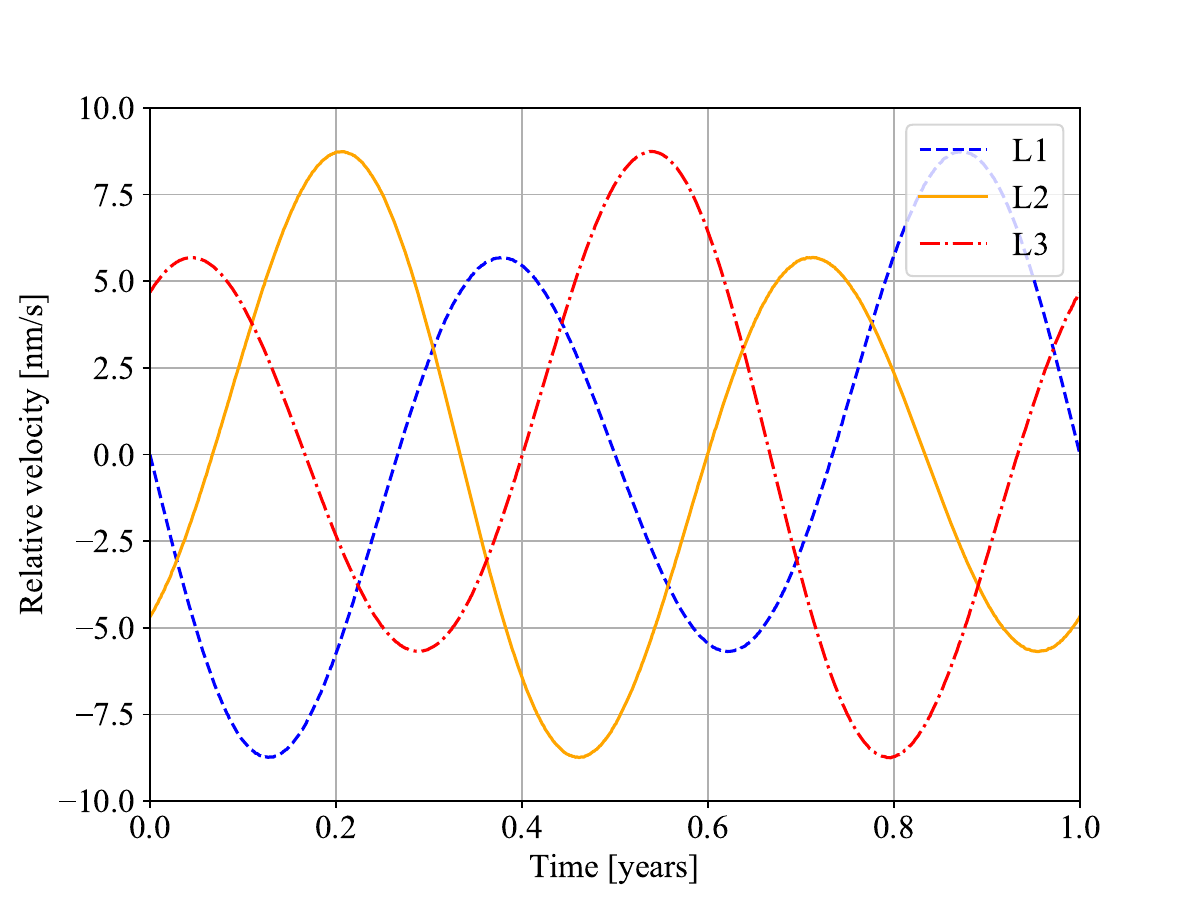}
    \caption{Variations in the relative velocities of the arm lengths for a year.}
    \label{fig:relv}
\end{figure}

Figure \ref{fig:relv} shows the result of the numerical simulation for the arm length variations for a nominal arm length of $100\,\mathrm{km}$. It shows that the maximum drift velocity can be on the order of $10\,\mathrm{nm/s}$ at most. We note that the largest velocity can be reduced by tuning the tilt angle. Therefore, we set the values of $v_2$ and $v_3$ to be on the order of $10\,\mathrm{nm/s}$ in the simulation in section \ref{sec:sensitivity}.

%text

\begin{acknowledgments}
We thank K.~Yamamoto, Y.~Okuma, R.~Sugimoto, T.~Akutsu, K.~Komori, Y.~Michimura, M.~Ando, K.~McKenzie, A.~Wade, and E.~Rees for fruitful discussions.
This work was supported by JSPS KAKENHI Grant No. 24K00651.
S.~S. was supported by FoPM, WINGS Program, the University of Tokyo.
\end{acknowledgments}

%\bibliographystyle{./unsrt_customized}

%\bibliography{./MyLibrary}

\begin{thebibliography}{10}

\bibitem{abbott2016b}
B.~P. Abbott {\em et~al.} , {\em Observation of {{Gravitational Waves}} from a {{Binary Black Hole Merger}}}.
\newblock Physical Review Letters, {\bfseries 116}(6) 061102, (2016).

\bibitem{abbott2017d}
B.~P. Abbott {\em et~al.} , {\em {{GW170817}}: {{Observation}} of {{Gravitational Waves}} from a {{Binary Neutron Star Inspiral}}}.
\newblock Physical Review Letters, {\bfseries 119}(16) 161101, (2017).

\bibitem{amaro-seoane2017}
P.~{Amaro-Seoane}, {\em Laser {{Interferometer Space Antenna}}}, arXiv:1702.00786v3, (2017).

\bibitem{luo2016}
J.~Luo {\em et~al.} , {\em {{TianQin}}: A space-borne gravitational wave detector}.
\newblock Classical and Quantum Gravity, {\bfseries 33}(3) 035010, (2016).

\bibitem{hu2017a}
W.-R. Hu and Y.-L. Wu, {\em The {{Taiji Program}} in {{Space}} for gravitational wave physics and the nature of gravity}.
\newblock National Science Review, {\bfseries 4}(5) 685--686, (2017).

\bibitem{kawamura2006}
S.~Kawamura {\em et~al.} , {\em The {{Japanese}} space gravitational wave antenna---{{DECIGO}}}.
\newblock Classical and Quantum Gravity, {\bfseries 23}(8) S125--S131, (2006).

\bibitem{harry2006}
G.~M. Harry, P.~Fritschel, D.~A. Shaddock, W.~Folkner, and E.~S. Phinney, {\em Laser interferometry for the {{Big Bang Observer}}}.
\newblock Classical and Quantum Gravity, {\bfseries 23}(15) 4887, (2006).

\bibitem{kuns2020}
K.~A. Kuns, H.~Yu, Y.~Chen, and R.~X. Adhikari, {\em Astrophysics and cosmology with a decihertz gravitational-wave detector: {{TianGO}}}.
\newblock Physical Review D, {\bfseries 102}(4) 043001, (2020).

\bibitem{izumi2021a}
K.~Izumi and M.-K. Fujimoto, {\em A back-linked {{Fabry}}--{{P\'erot}} interferometer for space-borne gravitational wave observations}.
\newblock Progress of Theoretical and Experimental Physics, {\bfseries 2021}(7) 073F01, (2021).

\bibitem{yamamoto2023}
K.~Yamamoto, {\em Intersatellite Clock Synchronization and Absolute Ranging for Gravitational Wave Detection in Space}.
\newblock Ph.D. thesis, Leibniz Universit\"at Hannover, (2023).

\bibitem{hellings2001}
R.~W. Hellings, {\em Elimination of clock jitter noise in spaceborne laser interferometers}.
\newblock Physical Review D, {\bfseries 64}(2) 022002, (2001).

\bibitem{tinto2018}
M.~Tinto and O.~Hartwig, {\em Time-delay interferometry and clock-noise calibration}.
\newblock Physical Review D, {\bfseries 98}(4) 042003, (2018).

\bibitem{hartwig2021}
O.~Hartwig and J.-B. Bayle, {\em Clock-jitter reduction in {{LISA}} time-delay interferometry combinations}.
\newblock Physical Review D, {\bfseries 103}(12) 123027, (2021).

\bibitem{yamamoto2022}
K.~Yamamoto {\em et~al.} , {\em Experimental verification of intersatellite clock synchronization at {{LISA}} performance levels}.
\newblock Physical Review D, {\bfseries 105}(4) 042009, (2022).

\bibitem{xie2023}
S.~Xie {\em et~al.} , {\em Bi-directional PRN laser ranging and clock synchronization for TianQin mission}.
\newblock Optics Communications, {\bfseries 541} 129558, (2023).

\bibitem{zeng2023}
H.~Zeng {\em et~al.} , {\em Experimental demonstration of weak-light inter-spacecraft clock jitter readout for TianQin}.
\newblock Optics Express, {\bfseries 31}(21) 34648, (2023).

\bibitem{xu2024}
M.-Y. Xu, Y.-J. Tan, and C.-G. Shao, {\em Clock-jitter noise reduction by sideband arm locking for space-borne gravitational wave detectors}.
\newblock Physical Review D, {\bfseries 110}(10) 102003, (2024).

\bibitem{xia2025}
Y.~Xia, A.~Fang, M.~Xu, Y.~Tan, and C.~Shao, {\em Clock {{Noise Suppression Techniques}} in {{Space-Borne Gravitational Wave Detection}}: {{A Review}}}.
\newblock Symmetry, {\bfseries 17}(8) 1314, (2025).

\bibitem{nakamura2016}
T.~Nakamura {\em et~al.} , {\em Pre-{{DECIGO}} can get the smoking gun to decide the astrophysical or cosmological origin of {{GW150914-like}} binary black holes}.
\newblock Progress of Theoretical and Experimental Physics, {\bfseries 2016}(9) 093E01, (2016).

\bibitem{lilley2021}
M.~Lilley {\em et~al.} , {\em {{ACES}}/{{PHARAO}}: High-performance space-to-ground and ground-to-ground clock comparison for fundamental physics}.
\newblock GPS Solutions, {\bfseries 25} 34, (2021).

\bibitem{bode2024}
C.~H. Bode, {\em Noise in the {{LISA}} Phasemeter}.
\newblock Ph.D. thesis, Leibniz Universit\"at Hannover, (2024).

\bibitem{kokuyama2016}
W.~Kokuyama, H.~Nozato, A.~Ohta, and K.~Hattori, {\em Simple digital phase-measuring algorithm for low-noise heterodyne interferometry}.
\newblock Measurement Science and Technology, {\bfseries 27}(8) 085001, (2016).

\bibitem{anai2024}
K.~Anai {\em et~al.} , {\em Quantum-enhanced optical phase-insensitive heterodyne detection beyond 3-{{dB}} noise penalty of image band}.
\newblock Optics Express, {\bfseries 32}(11) 19372--19387, (2024).

\bibitem{chalermsongsak2015}
T.~Chalermsongsak {\em et~al.} , {\em Broadband measurement of coating thermal noise in rigid {{Fabry}}--{{P\'erot}} cavities}.
\newblock Metrologia, {\bfseries 52}(1) 17--30, (2015).

\bibitem{gras2018}
S.~Gras and M.~Evans, {\em Direct measurement of coating thermal noise in optical resonators}.
\newblock Physical Review D, {\bfseries 98}(12) 122001, (2018).

\bibitem{eisele2009}
{\relax Ch}.~Eisele, A.~{\relax Yu}. Nevsky, and S.~Schiller, {\em Laboratory {{Test}} of the {{Isotropy}} of {{Light Propagation}} at the $\mathit{10^{-17}}$ {{Level}}}.
\newblock Physical Review Letters, {\bfseries 103}(9) 090401, (2009).

\bibitem{nayak2006}
K.~R. Nayak, S.~Koshti, S.~V. Dhurandhar, and J.-Y. Vinet, {\em On the minimum flexing of {{LISA}}'s arms}.
\newblock Classical and Quantum Gravity, {\bfseries 23}(5) 1763, (2006).

\end{thebibliography}
%\input{clock_prd1.bbl}

\end{document}